\documentclass[conference]{IEEEtran}

\usepackage{subcaption}
\usepackage[pdftex]{graphicx}
\usepackage[export]{adjustbox}
\usepackage{url}
\usepackage[nospace]{cite}
\usepackage{ifthen}

\usepackage{tikz}

\newcommand\submittedtext{%
  \footnotesize This work has been submitted to the IEEE for possible publication. Copyright may be transferred without notice, after which this version may no longer be accessible.}

\newcommand\submittednotice{%
\begin{tikzpicture}[remember picture,overlay]
\node[anchor=south,yshift=10pt] at (current page.south) {\fbox{\parbox{\dimexpr0.65\textwidth-\fboxsep-\fboxrule\relax}{\submittedtext}}};
\end{tikzpicture}%
}

\begin{document}

\title{An Educational Tool for Learning about Social Media Tracking, Profiling, and Recommendation}
\def\screenshotsize{0.6} 

%

\newboolean{shownames}
\setboolean{shownames}{true}

\ifthenelse{\boolean{shownames}} {
\author{\IEEEauthorblockN{Nicolas Pope\IEEEauthorrefmark{1},
Juho Kahila\IEEEauthorrefmark{2},
Jari Laru\IEEEauthorrefmark{3}, 
Henriikka Vartiainen\IEEEauthorrefmark{2},
Teemu Roos\IEEEauthorrefmark{4} and
Matti Tedre\IEEEauthorrefmark{1}}
\IEEEauthorblockA{\IEEEauthorrefmark{1}School of Computing, 
University of Eastern Finland,
Joensuu, Finland\\ Email: firstname.lastname@uef.fi}
\IEEEauthorblockA{\IEEEauthorrefmark{2}School of Applied Educational Science and Teacher Education, 
University of Eastern Finland,
Joensuu, Finland\\ Email: firstname.lastname@uef.fi}
\IEEEauthorblockA{\IEEEauthorrefmark{3}University of Oulu, Faculty of Education and Psychology\\
Email: firstname.lastname@oulu.fi}
\IEEEauthorblockA{\IEEEauthorrefmark{4}Department of Computer Science, 
University of Helsinki,
Finland\\ Email: firstname.lastname@helsinki.fi}}
}
{
\author{\IEEEauthorblockN{~
}
\IEEEauthorblockA{~\\~\\ 
}
\IEEEauthorblockA{~\\~\\ 
}
\IEEEauthorblockA{~\\~\\ 
}
\IEEEauthorblockA{~\\~\\ 
}
}
}


\maketitle

\submittednotice

\begin{abstract}
This paper introduces an educational tool for classroom use, based on explainable AI (XAI), designed to demystify key social media mechanisms---tracking, profiling, and content recommendation---for novice learners.  The tool provides a familiar, interactive interface that resonates with learners' experiences with popular social media platforms, while also offering the means to ``peek under the hood'' and exposing basic mechanisms of datafication. Learners gain first-hand experience of how even the slightest actions, such as pausing to view content, are captured and recorded in their digital footprint, and further distilled into a personal profile.  The tool uses real-time visualizations and verbal explanations to create a sense of immediacy: each time the user acts, the resulting changes in their engagement history and their profile are displayed in a visually engaging and understandable manner.  
This paper discusses the potential of XAI and educational technology in transforming data and digital literacy education and in fostering the growth of children's privacy and security mindsets.
\end{abstract}

\IEEEpeerreviewmaketitle

\section{Introduction}


Social media has become an integral element of daily life for many people, fundamentally influencing the way people of all ages, including children, perceive and interact with the world \cite{stoilova20}.  Its impact has been the focus of sustained research efforts, and educational interventions have followed.  Research studies have highlighted the importance of educating users, especially the youth, about how to mitigate risks related to social media, focusing on misinformation, privacy, and personal cybersecurity \cite{livingstone:2014,selwyn18,pangrazio19,stoilova20,keen22}.  Consequently, social media education has primarily meant guidelines on themes like information literacy, privacy policies, and preventing cyberbullying \cite{swart:2021,halama:2022,ng:2022}.  Many of these initiatives avoid discussing the actual mechanisms and algorithms used by social media platforms, such as profiling, so that most people resort to ``folk theories'' when asked to explain their functioning~\cite{eslami:2016,buchi:2023}. 

We argue that, while important, education focusing on personal safety awareness and guidelines is not enough: learning about the mechanisms underlying social media platforms is necessary for preparing learners to understand the algorithm-driven dynamics on social media platforms, including ``how echo chambers are formed, emotions amplified, and behavior engineered---and through them how products are sold, elections swayed, and mass movements born'' \cite{valtonen19}. Going beyond surface-level understanding also enables children to assume a more active role in envisioning alternative solutions that better meet their needs and expectations.


A major challenge in educating young learners about the mechanisms of social media lies in the absence of practical, classroom-ready tools that work at a level of abstraction that exposes the mechanisms at play but does not bury those mechanisms under complicated technical detail.  Unplugged methods can provide a bird's eye view on demystifying these complex mechanisms, but they fall short in terms of demonstrating how automation and scale change the dynamics.  Unplugged methods are also challenged by the multiplicity of ways in which behavioral patterns are captured and aggregated into digital footprints and targeted content through profiling and recommendation algorithms.  

The gap between learning material and reality is particularly concerning in light of growing concerns around the ethical implications of tracking, profiling, and recommendation systems. 
The need for innovative educational tools for filling this gap is evident---we need tools elucidating mechanisms underlying social media platforms in safe and engaging ways, allowing children to learn through hands-on experience without compromising their personal data.  

\begin{figure}[!t]
\centering
\includegraphics[width=0.99\linewidth, frame]{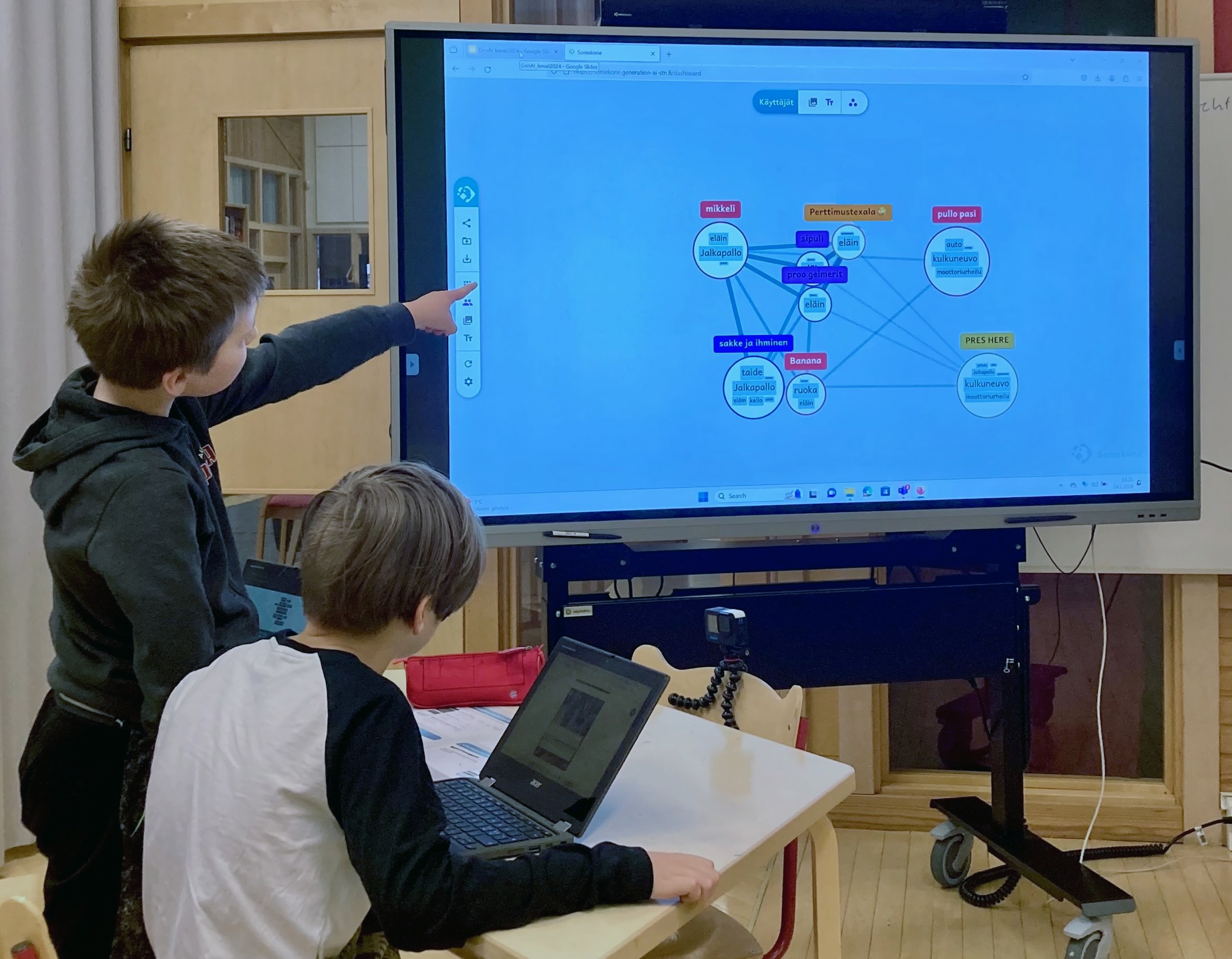}
\caption{Children using Somekone on two devices: One for browsing the feed, one for analytics of the other.  The classroom view shows a social network of the classroom.}
\label{fig:graph_photo}
\end{figure}

This paper presents Somekone (\url{https://somekone.generation-ai-stn.fi/}), an educational tool for learning the mechanisms of data collection, profiling, and recommendation in social media.  The tool provides learners an image feed app with the usual social media engagement modalities: like, follow, share, comment, and react with emoji.  Learners can monitor and explore the depth of data collection either on the app or by connecting another smartphone for monitoring.  They can explore how profiles develop through different kinds of user engagement, and explore how recommendations are made based on data and profiles.  At the same time the teacher can visualize on a projector how the classroom's social network develops in real-time as learners scroll through images and engage with them.  Fig. \ref{fig:graph_photo} shows a photo from a classroom intervention, with two laptops paired to show an image feed and analytics data, as well as the social network display for the whole classroom.

This paper presents a technological proof of concept on teaching children about the inner workings of social media platforms.  It presents the design principles, security and privacy features, and functionalities of Somekone.  It demonstrates an approach to teaching children the principles of algorithm-driven automation on social media platforms, done at an abstraction level that does not require advanced mathematical concepts or programming skills.  It advocates an approach that relies on immediacy of feedback between learners' browsing activities and analytics views.  The tool we describe serves as an example of GDPR-safe social media educational technology that respects children's privacy in a safe learning environment.  The paper is aimed at educators, researchers, and policymakers interested in developing social media education in K--12 schools.

\section{Design of the Tool}

Somekone (literally translated as ``\underline{so}cial \underline{me}dia machine'') has a user interface reminiscent of Instagram, which allows children to learn in an easy-to-use environment that many are already familiar with \cite{vartiainen23c}. It provides an infinite scrolling feed of images, coupled with a set of generic social media engagement features, such as liking, reacting with emoji, commenting, following other users, and sharing with everyone, friends, or privately.  The image set is derived from a hand-picked and labeled set of 727 images, selected from Pixabay by two classes of eight-graders and labeled by the same children at the time of selection.  The dataset was further curated by two researchers.  Uploading children's own images is prohibited to prevent any uncurated, age-restricted material from entering the classroom, and to maintain image tag cohesion.


The tool is designed to respect children's privacy.  All engagement data are collected and  stored locally within the teacher's laptop in the classroom; none of children's data are stored on external servers.  Only the app, along with the image and label dataset, are retrieved from outside the classroom.  The tool uses WebRTC\footnote{\url{https://webrtc.org/}} for peer-to-peer communication between devices, eliminating the need for an Internet connection beyond the initial signaling, provided the devices are in the same network.


\begin{figure}[!t]
\centering
\includegraphics[width=\screenshotsize\linewidth, frame]{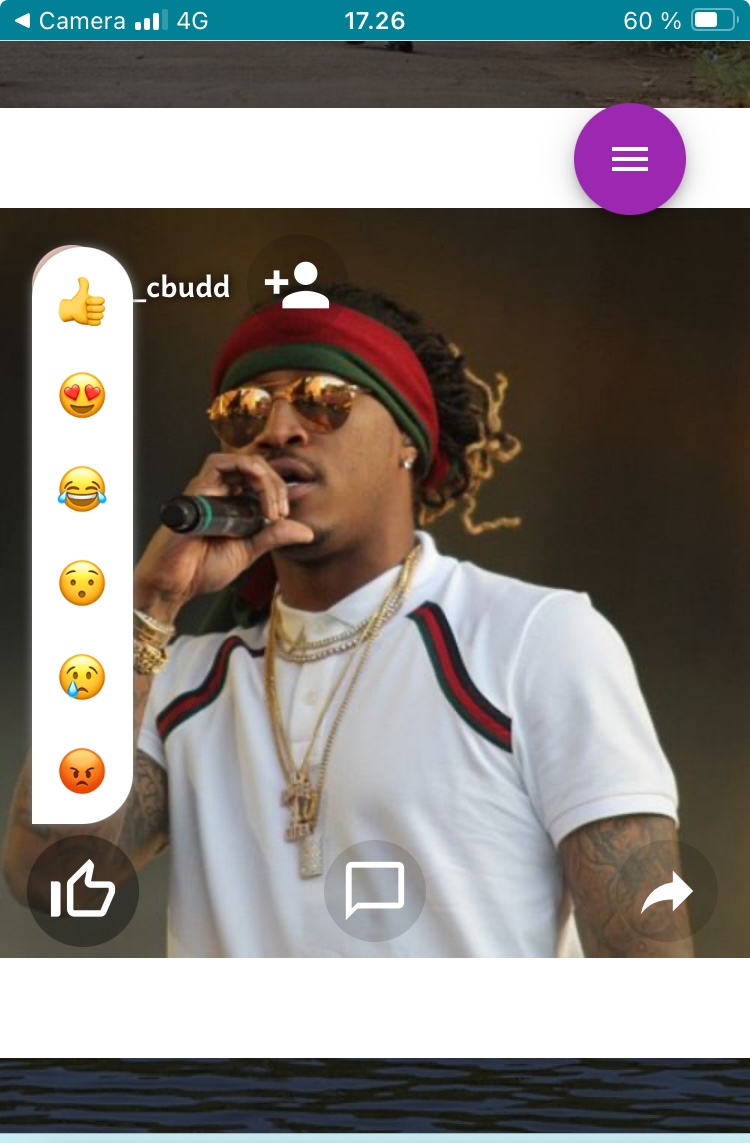}
\caption{A screenshot of Jarmo browsing images on Somekone on his phone, with ``like'' button showing the range of available emojis.}
\label{fig:app_browsing}
\end{figure}

\def\minipicturesize{0.9} 

\begin{figure*}[!t]
    \centering
    \begin{subfigure}[t]{0.33\textwidth}
        \centering
        \captionsetup{justification=centering}
        \includegraphics[width=\minipicturesize\textwidth, frame]{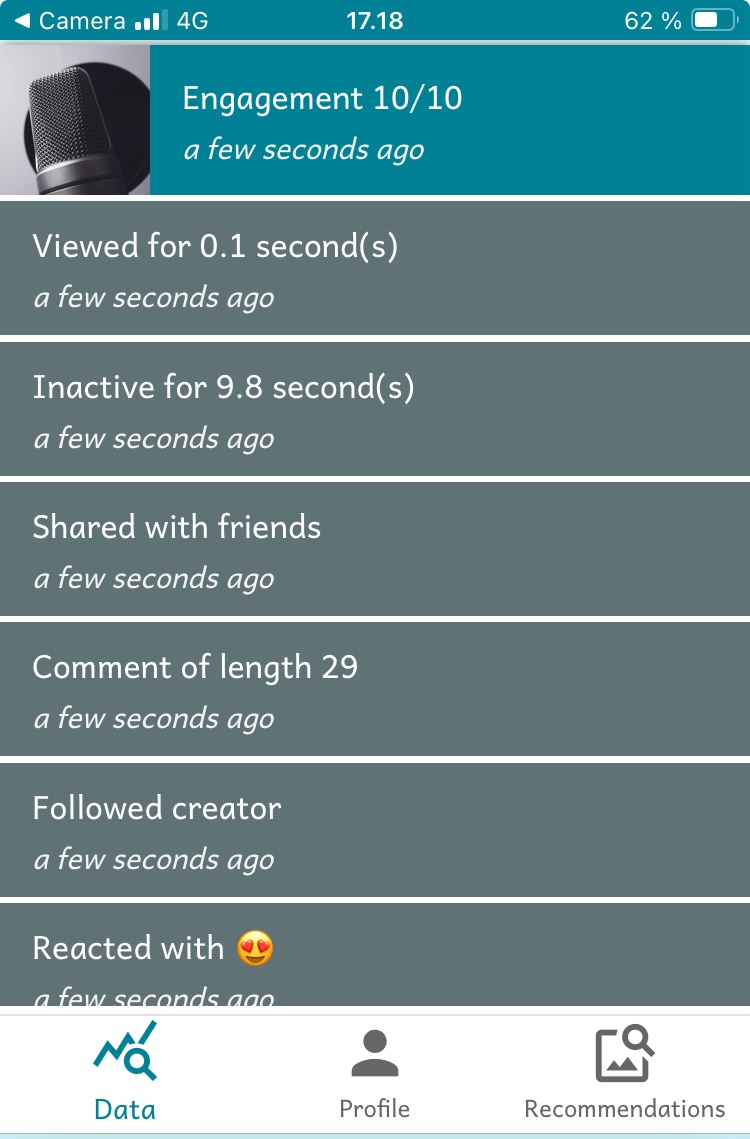}
        \caption{All generated browsing data.\\(Anna's phone)}
        \label{fig:app_dataview}
    \end{subfigure}%
    ~ 
    \begin{subfigure}[t]{0.33\textwidth}
        \centering
        \captionsetup{justification=centering}
        \includegraphics[width=\minipicturesize\linewidth, frame]{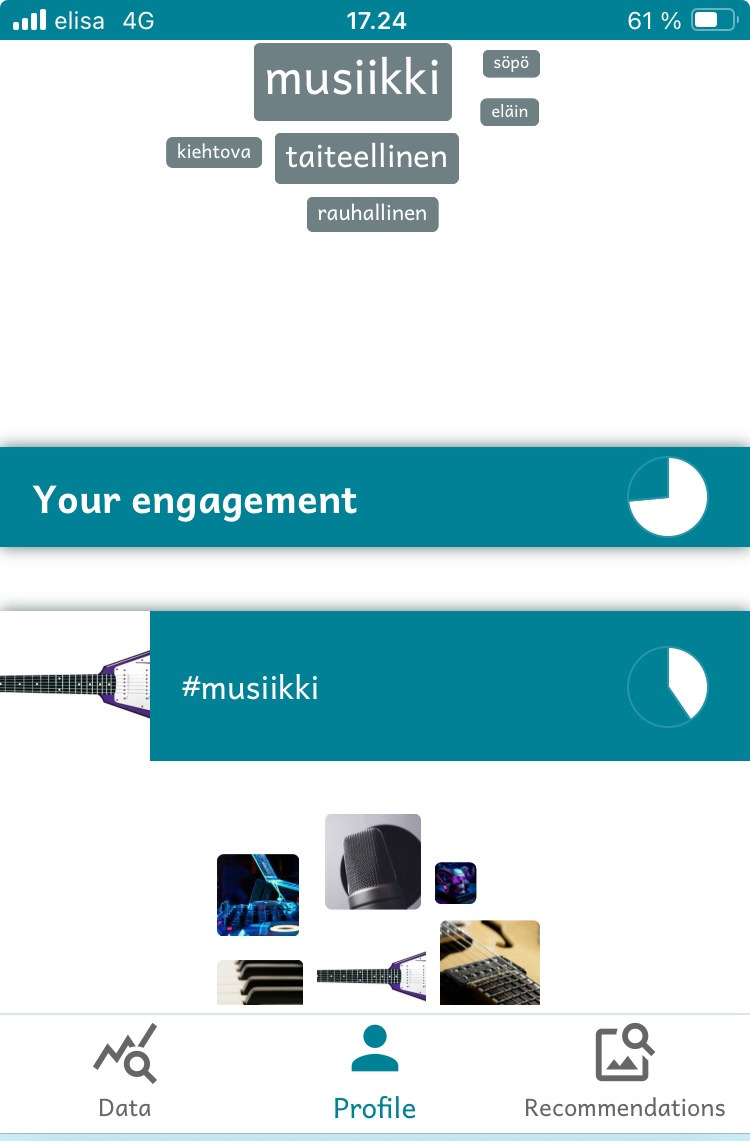}
        \caption{Somekone-generated profile of Jarmo.\\(Anna's phone)}
        \label{fig:app_profile}
    \end{subfigure}%
    ~ 
    \begin{subfigure}[t]{0.33\textwidth}
        \centering
        \captionsetup{justification=centering}
        \includegraphics[width=\minipicturesize\linewidth, frame]{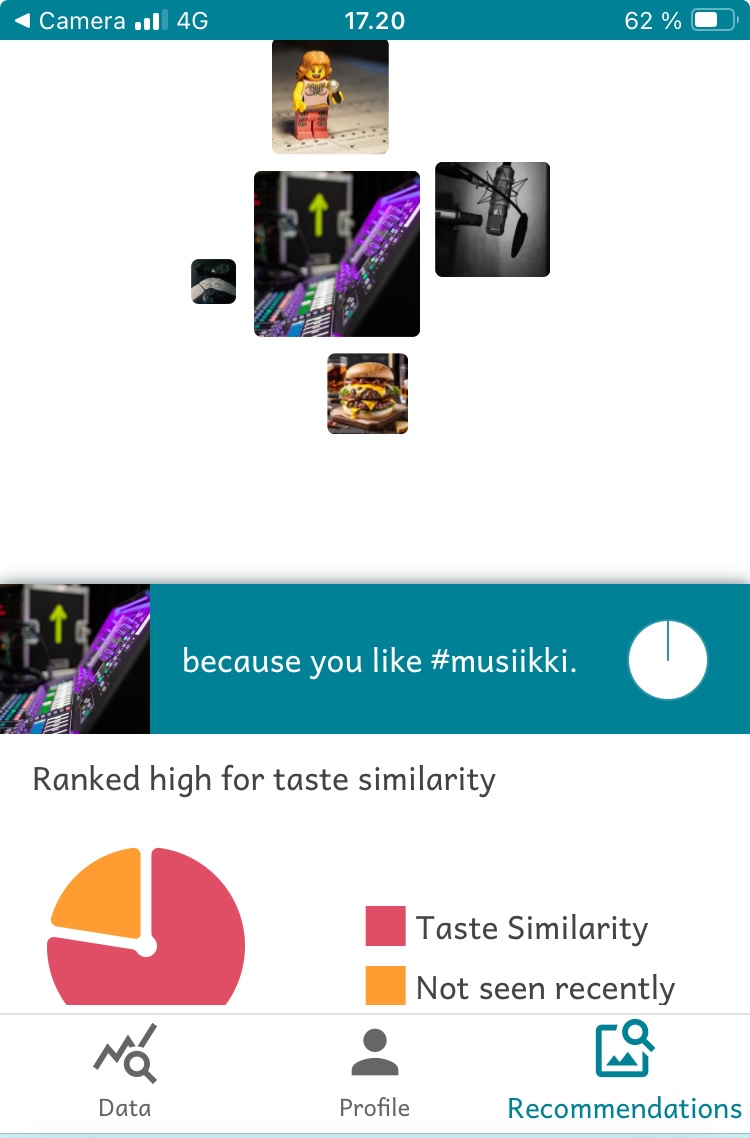}
        \caption{Upcoming recommendations, explained.\\(Anna's phone)}
        \label{fig:app_recommender}
    \end{subfigure}
    \caption{Anna is Jarmo's pair in the exercise, and her device is connected to Jarmo's device.  Anna can choose between real-time views of all the data Jarmo's browsing generates, the profile Somekone has built of Jarmo, and the upcoming recommendations and verbal explanations for them (``because you liked...'', ``ranked high for taste similarity...'', etc).}
\end{figure*}

The three key mechanisms that Somekone is designed to illustrate are $(i)$ tracking, $(ii)$ profiling, and $(iii)$ content recommendation. Additionally, it provides a detailed view into $(iv)$ the social network formed by building a graph where users with similar profiles are grouped near each other. Below we describe these key concepts and functionalities.

\subsection{Data Collection (Tracking)}
The first of the three key social media mechanisms that Somekone is designed to teach is data collection (tracking). It is implemented by tracking a variety of engagement metrics from the user's browsing session.  These metrics include seeing an image, time spent viewing an image, likes, comments and their length, periods of user inactivity, reactions with one of five different emoji, following the image creator, and sharing the image privately, with friends, or publicly.  Given the classroom context, some data collection channels, such as location data, cannot be captured. Consequently, the typical Instagram features like likes, follows, and comments are slightly extended to include additional reaction types found on other platforms.  Figure \ref{fig:app_browsing} shows a screenshot of the interface when the user engages with the like button, revealing the range of available emoji reactions.




To facilitate exploration of the relationship between their engagement scores and the data collected in the user's action log during their browsing sessions, children can access a real-time visualization of the action log.   Somekone offers two methods for this: An integrated ``Engagement Overview'' tab for individual use case, and the same view from a separate device for pair work (Figure \ref{fig:app_dataview}).  This paper presents a pair work example: ``Jarmo'' browses the image feed on his device and ``Anna'' monitors, on her own device, the underlying data collection process as Jarmo engages with the image feed.  In Figure \ref{fig:app_dataview} Anna connects her device to Jarmo's, allowing her to view a live feed of all the data Somekone is tracking about Jarmo's session.  For example, her feed indicates Jarmo's high level of engagement with a microphone image (10/10 engagement), including sharing it with friends, commenting, following its creator, and reacting with a ``heart eyes'' emoji.  As Jarmo continues to browse his feed, Anna's device displays the collected data in real time.


\subsection{Profiling (Modeling)}
The second key concept that Somekone is designed to teach is profiling.  This is implemented by transforming user actions into profiles, clustering them, and visualizing the results. The profiles in this tool are primarily composed of topic affinities, updated based on the user's engagement with labeled images. The user's action log is then used to generate an engagement score for each image. Each type of interaction carries a different weight, contributing more or less to the engagement score. These scores, combined with the labels of the images, update the user's affinity for each label. The set of label affinities is transformed into a taste vector, which can be compared with other users using cosine similarity (see, e.g.,~\cite{tan:2005}). The resulting similarity score is used as the weight on the social network graph's edges, facilitating clustering of users with similar tastes in the visualization.  To enhance recommendation accuracy, each user's profile includes a set of learned weights indicating the most effective recommendation strategies or features for that user. More complex profiling, such as demographic identification or using features beyond topic affinities, is not pursued due to limited resources and data in the browser, and due to the increasing difficulty to explain such features to learners.

\begin{figure*}[!t]
\centering
\includegraphics[width=0.99\linewidth, frame]{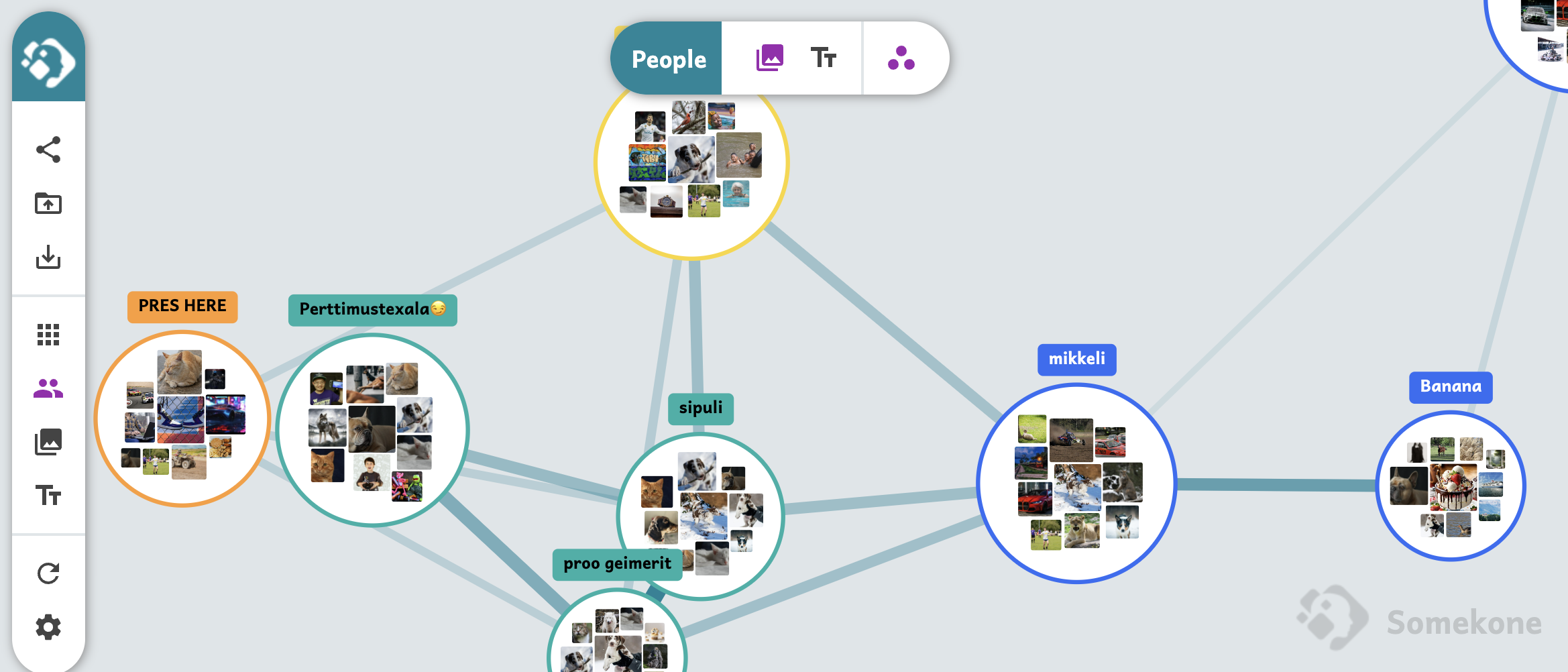}
\caption{Social network of children browsing Somekone in the classroom (displayed on the classroom projector).  Thick lines between nodes indicate a strong similarity in taste profile vectors.  Nodes are clustered by the similarity of topic (tag) engagement, as indicated by their proximity and node color.}
\label{fig:image_network}
\end{figure*}

\begin{figure*}[!t]
    \centering
    \begin{subfigure}[t]{0.33\textwidth}
        \centering
        \includegraphics[width=\linewidth, frame]{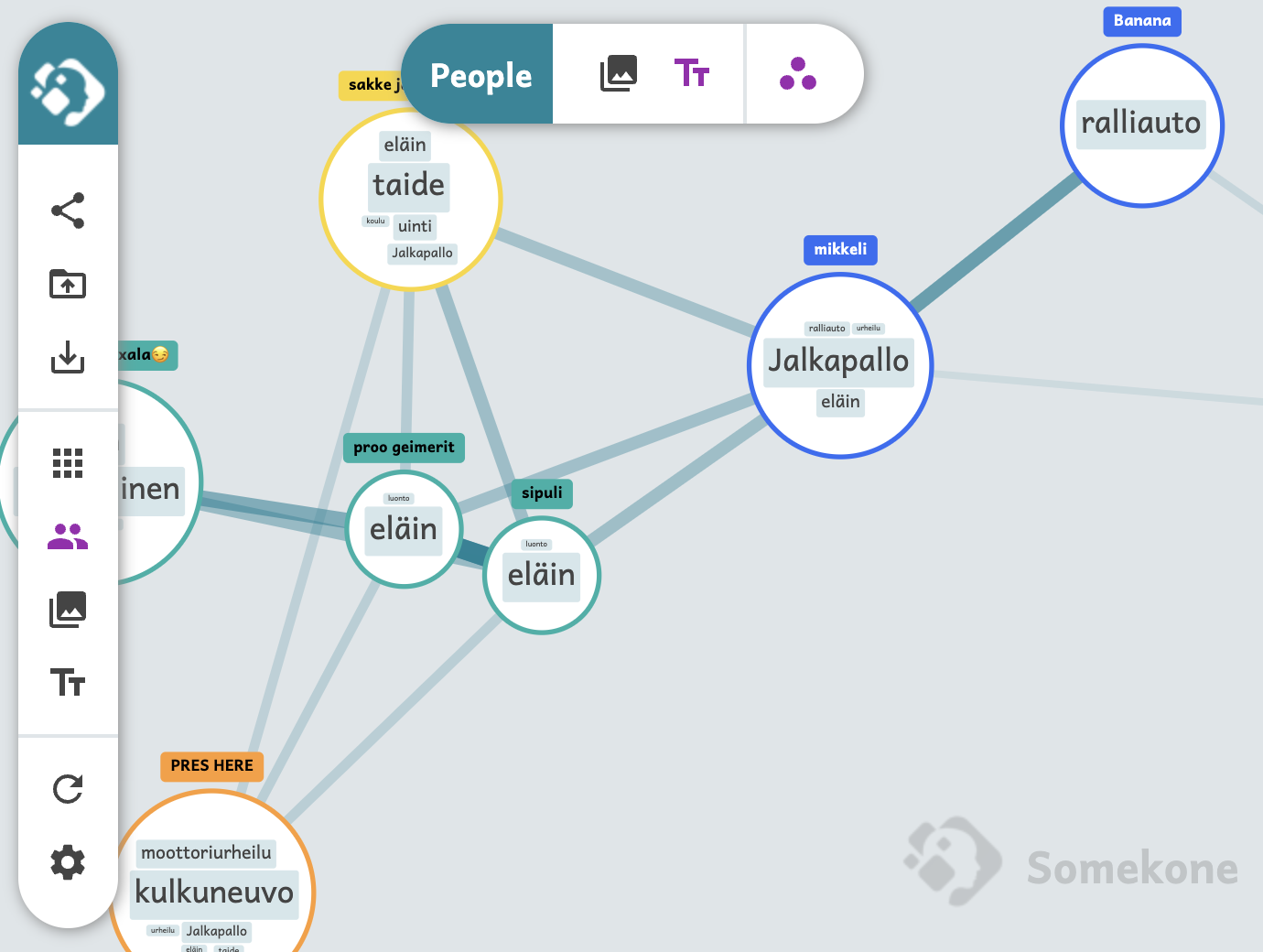}
        \caption{Children and their most engaged topics.}
        \label{fig:mini_users}
    \end{subfigure}%
    ~ 
    \begin{subfigure}[t]{0.33\textwidth}
        \centering
        \includegraphics[width=\linewidth, frame]{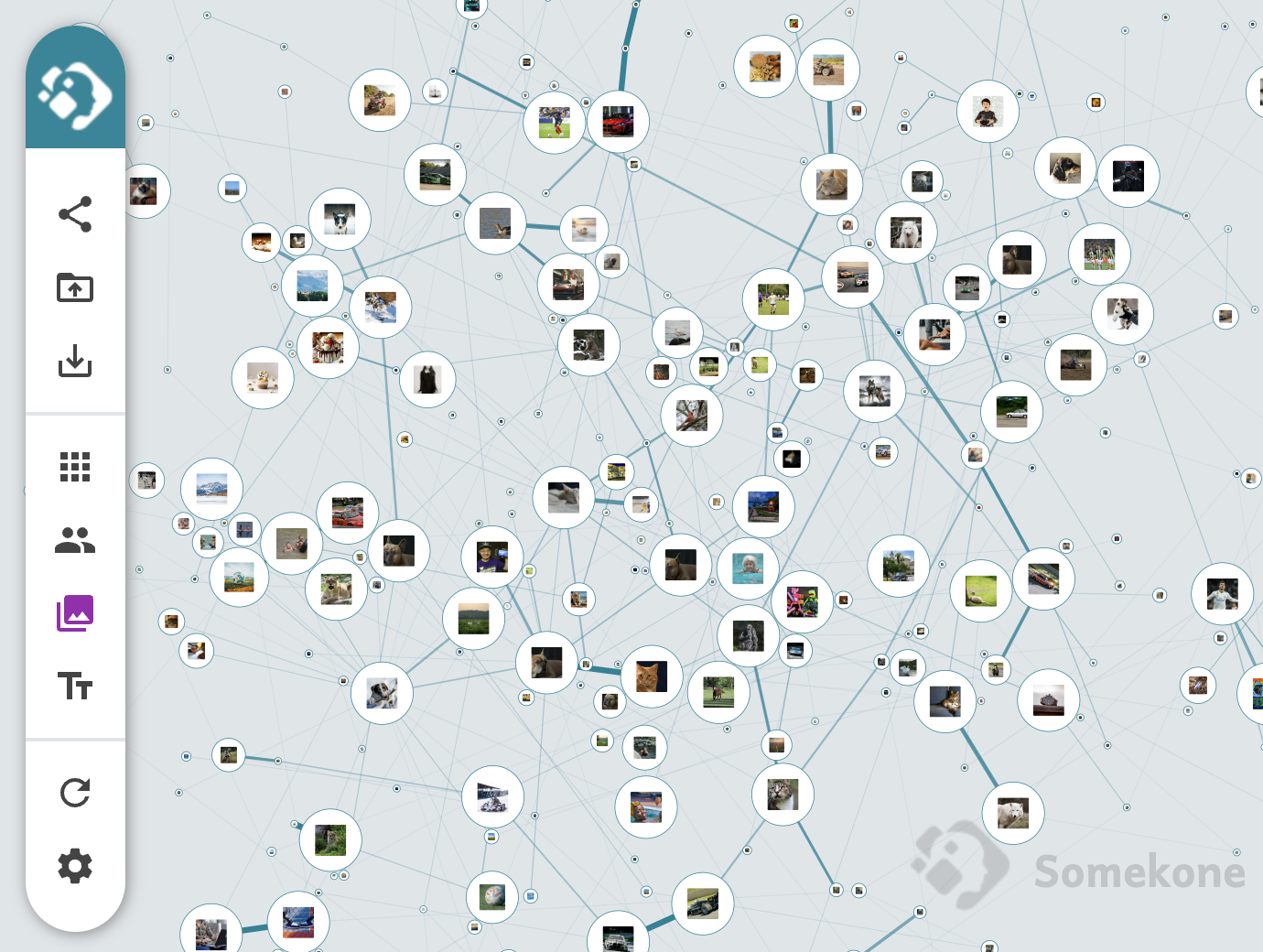}
        \caption{Picture co-engagement network.}
        \label{fig:mini_pics}
    \end{subfigure}%
    ~ 
    \begin{subfigure}[t]{0.33\textwidth}
        \centering
        \includegraphics[width=\linewidth, frame]{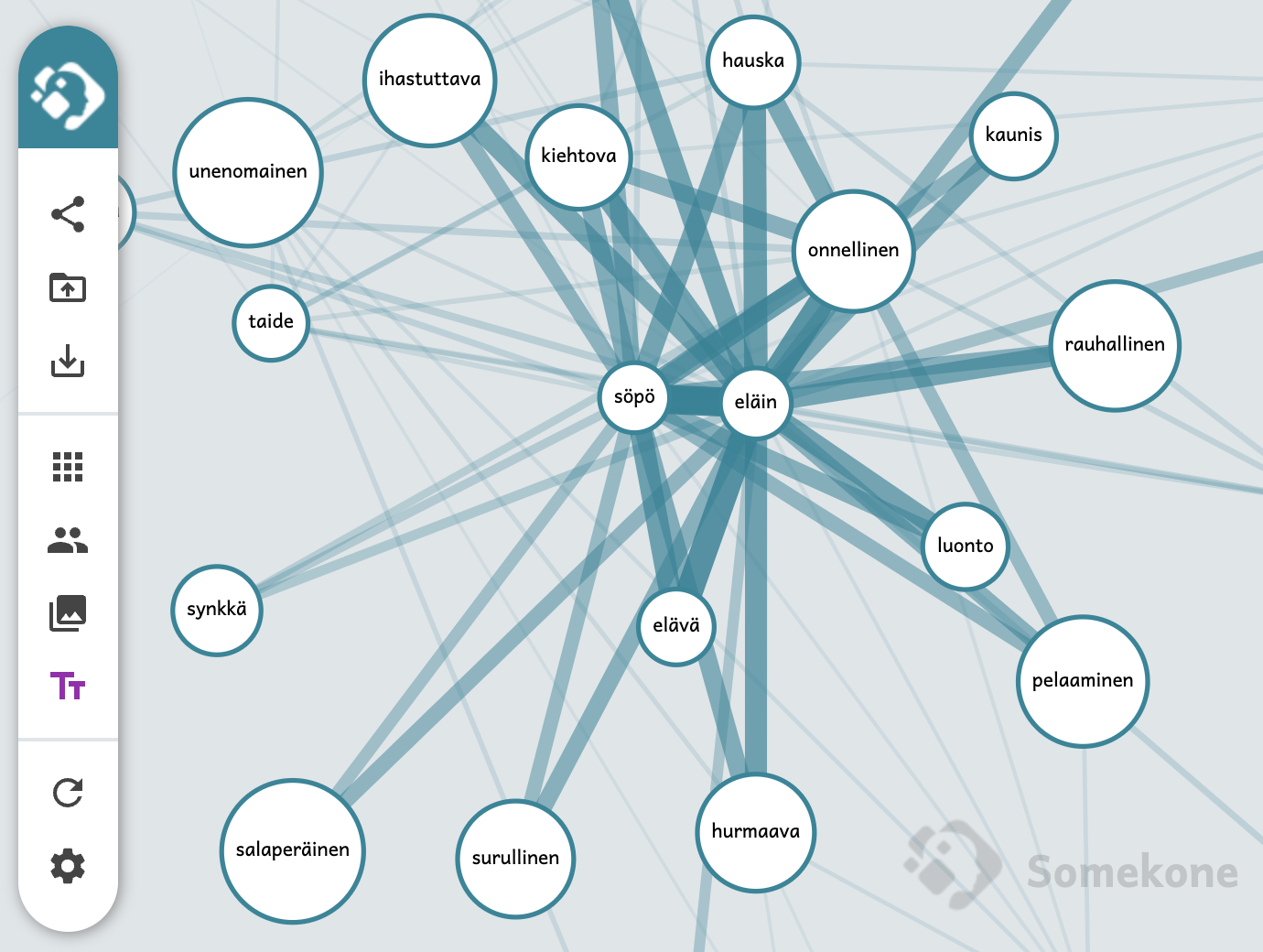}
        \caption{Topic co-engagement network.}
        \label{fig:mini_topics}
    \end{subfigure}
    \caption{Three example views of the data (displayed on the classroom projector)}
\end{figure*}

Content profiling is also enhanced by user data, where related images are connected using co-engagements.  This method of collaborative filtering allows for the identification of relationships between images beyond their labels, although it requires substantial interaction data.  Figure \ref{fig:mini_pics} presents a graph of image co-engagements after a learning session with 18 learners.


To facilitate children's exploration of how their profiles are created, Somekone offers a range of visualizations of the profiling mechanisms. The tool combines image tags and engagement scores to create a list of the most engaged topics, visualizing those tags as a word cloud.  This word cloud is broken down into individual tags, and for each tag the tool displays a set of images that the user engaged with more than others.  It also visualizes which topics received the most engagement through actions such as sharing, following, reacting with emoji, and viewing.  Figure \ref{fig:app_profile} shows a screenshot where Anna is analyzing Jarmo's profile on her device.  The profile is updated in real-time as Jarmo continues to interact with the image feed.  Anna sees that Jarmo's most engaged topics are \#musiikki (music) and \#taiteellinen (artistic), and by scrolling further, she can view a breakdown of the most engaged \#musiikki-tagged images and other visualizations of Jarmo's profile.


\subsection{Recommendation}

The third key concept to be taught, recommendation, was implemented by ranking image candidates based on a number of criteria (for an overview of recommendation systems, see, e.g.,~\cite{ricci:2011}): a) collaborative-filtering using engagement by other similar user profiles; b) content-based filtering using the user's topic affinities identified from tag engagements; c) collaborative-filtering via image co-engagement and popularity (images engaged with together in a session); and d) occasionally, random selection. Each of these candidates is scored and ranked using a range of user and content features, with the most significant score component being provided as an explanation of the ranking.  Children learn that recommendations in social media do not require the systems to \textit{understand} or care about the image content or semantics.  Tags are not essential, either, although they improve the quality of recommendations.  The key requirement is engagement data from numerous users.  They also discover that their continuous interaction with social media not only enhances their personal experience but also influences other users' experience with the platform, too.  

Figure \ref{fig:app_recommender} shows a screenshot where Anna is analyzing Jarmo's recommendations.  She can see Jarmo's queue of the next images and scroll down for a detailed explanation of why each of the next five images has been queued.  For example, in Fig.\ \ref{fig:app_recommender}, the next recommended image is a photo of studio mixing table with neon colors, and Somekone provides a verbal explanation saying it was selected due to its high taste similarity to Jarmo's most engaged topic \#musiikki, and because it had not appeared in Jarmo's feed recently.


\subsection{Social network}
To illustrate  what ``similarity'' means and to show how clusters of similar users develop in social media services, Somekone shows a real-time visualization of social network of the current class, clustered by profile similarity.  Children learn that clustering forms around not just their friends or whom they follow but around anyone with a similar profile.  They also learn how clustering influences the recommendations they see.  The visualization uses a force layout where edges act as springs and nodes are charged to repel each other. User similarity is used as the edge weight which determines the strength and length of the edge in the visualization, a strong similarity score between users results in shorter and stronger edges. Consequently, similar users move together to form clusters. Colouring is achieved by propagation along the similarity weighted edges.  

Figure \ref{fig:image_network} shows an example teacher view from classroom testing, shown to children on the classroom projector.  It shows eight children's profiles as nodes labeled by their nicknames and filled with their most engaged images.  Nodes are clustered by their profile similarity, based on tags in images they have engaged with the most.  The teacher can swap between a variety of visualizations such as tag cloud (Fig. \ref{fig:mini_users}) and engagement score, highlight profiles and their connections, examine each children's profile (data log, profile, recommendations), and display image co-engagement (Fig. \ref{fig:mini_pics}) or topic co-engagement (Fig. \ref{fig:mini_topics}) networks.





\section{Discussion}

This paper presents an educational tool designed to expose and explain the processes by which familiar social media platforms track users, profile them, and recommend content tailored to individual preferences.  It aims to foster the development of data agency and critical data-conscious mindsets among children at a crucial age when their data agency, ability to safely navigate the digital environment, and personal data strategies are evolving \cite{lupton17}.  It also aims to enable children to understand the implications of their online behavior to their social media experience.  

This tool bridges the gap in existing educational tools for teaching social media mechanisms by simplifying complex concepts without overwhelming young children with technical details.  The tool's main innovation lies in its real-time visualizations and the immediacy of its interactive elements, enhancing the learning experience: Every action a user takes on on one device immediately impacts the displays on other devices, illustrating the dynamics of data collection, profiling, and recommendations in an understandable and tangible manner.  The tool serves as an example of how explainable AI (XAI) principles can be applied in a practical, educational context to help learners develop their understanding of digital technologies beyond folk theories that people commonly resort to in order to explain their everyday experiences~\cite{eslami:2016}.

The tool supports experiential learning, encouraging children to learn through hands-on experience, and it encourages them to build their own understanding of social media mechanisms by experimenting with a system that exposes the underlying mechanisms of social media in a real-time visual environment.



\bibliographystyle{IEEEtran}
\bibliography{bibliography}

\begin{thebibliography}{10}
\providecommand{\url}[1]{#1}
\csname url@samestyle\endcsname
\providecommand{\newblock}{\relax}
\providecommand{\bibinfo}[2]{#2}
\providecommand{\BIBentrySTDinterwordspacing}{\spaceskip=0pt\relax}
\providecommand{\BIBentryALTinterwordstretchfactor}{4}
\providecommand{\BIBentryALTinterwordspacing}{\spaceskip=\fontdimen2\font plus
\BIBentryALTinterwordstretchfactor\fontdimen3\font minus \fontdimen4\font\relax}
\providecommand{\BIBforeignlanguage}[2]{{%
\expandafter\ifx\csname l@#1\endcsname\relax
\typeout{** WARNING: IEEEtran.bst: No hyphenation pattern has been}%
\typeout{** loaded for the language `#1'. Using the pattern for}%
\typeout{** the default language instead.}%
\else
\language=\csname l@#1\endcsname
\fi
#2}}
\providecommand{\BIBdecl}{\relax}
\BIBdecl

\bibitem{stoilova20}
M.~Stoilova, S.~Livingstone, and R.~Nandagiri, ``Digital by default: Children's capacity to understand and manage online data and privacy,'' \emph{Media and Communication}, vol.~8, no.~4, pp. 197--207, 2020.

\bibitem{livingstone:2014}
S.~Livingstone, ``Developing social media literacy: How children learn to interpret risky opportunities on social network sites,'' \emph{Communications}, vol.~39, no.~3, pp. 283--303, 2014.

\bibitem{selwyn18}
\BIBentryALTinterwordspacing
N.~Selwyn and L.~Pangrazio, ``Doing data differently? {D}eveloping personal data tactics and strategies amongst young mobile media users,'' \emph{Big Data \& Society}, vol.~5, no.~1, pp. 1--12, 2018. [Online]. Available: \url{https://doi.org/10.1177/2053951718765021}
\BIBentrySTDinterwordspacing

\bibitem{pangrazio19}
L.~Pangrazio and N.~Selwyn, ```{P}ersonal data literacies': A critical literacies approach to enhancing understandings of personal digital data,'' \emph{New Media \& Society}, vol.~21, no.~2, pp. 419--437, 2019.

\bibitem{keen22}
C.~Keen, ``Apathy, convenience or irrelevance? identifying conceptual barriers to safeguarding children's data privacy,'' \emph{New Media \& Society}, vol.~24, no.~1, pp. 50--69, 2022.

\bibitem{swart:2021}
J.~Swart, ``Experiencing algorithms: How young people understand, feel about, and engage with algorithmic news selection on social media,'' \emph{Social Media + Society}, vol.~7, no.~2, 2021.

\bibitem{halama:2022}
J.~Halama, T.~Frenzel, L.~Hofmann, C.~Klose, N.~Seifert, K.~Telega, and F.~Bocklisch, ``Is there a privacy paradox in digital social media use? the role of privacy concerns and social norms,'' \emph{Open Psychology}, vol.~4, no.~1, pp. 265--277, 2022.

\bibitem{ng:2022}
E.~D. Ng, J.~Y.~X. Chua, and S.~Shorey, ``The effectiveness of educational interventions on traditional bullying and cyberbullying among adolescents: A systematic review and meta-analysis,'' \emph{Trauma, Violence, \& Abuse}, vol.~23, no.~1, pp. 132--151, 2022.

\bibitem{eslami:2016}
M.~Eslami, K.~Karahalios, C.~Sandvig, K.~Vaccaro, A.~Rickman, K.~Hamilton, and A.~Kirlik, ``First {I} "like" it, then {I} hide it: Folk theories of social feeds,'' in \emph{Proceedings of the 2016 CHI Conference on Human Factors in Computing Systems}.\hskip 1em plus 0.5em minus 0.4em\relax Association for Computing Machinery, 2016, p. 2371–2382.

\bibitem{buchi:2023}
M.~B\"uchi, E.~Fosch-Villaronga, C.~Lutz, A.~Tam\`o-Larrieux, and S.~Velidi, ``Making sense of algorithmic profiling: user perceptions on facebook,'' \emph{Information, Communication \& Society}, vol.~26, no.~4, pp. 809--825, 2023.

\bibitem{valtonen19}
T.~Valtonen, M.~Tedre, K.~M\"akitalo, and H.~Vartiainen, ``Media literacy education in the age of machine learning,'' \emph{Journal of Media Literacy Education}, vol.~11, no.~1, pp. 20--36, 2019.

\bibitem{vartiainen23c}
H.~Vartiainen, J.~Kahila, M.~Tedre, E.~Sointu, and T.~Valtonen, ``More than fabricated news reports: Children's perspectives and experiences of fake news,'' \emph{Journal of Media Literacy Education}, vol.~15, no.~2, pp. 17--30, 2023.

\bibitem{tan:2005}
P.-N. Tan, M.~Steinbach, and V.~Kumar, \emph{Introduction to Data Mining}.\hskip 1em plus 0.5em minus 0.4em\relax USA: Addison-Wesley Longman Publishing Co., Inc., 2005.

\bibitem{ricci:2011}
F.~Ricci, L.~Rokach, and B.~Shapira, ``Introduction to recommender systems handbook,'' in \emph{Recommender Systems Handbook}, F.~Ricci, L.~Rokach, B.~Shapira, and P.~B. Kantor, Eds.\hskip 1em plus 0.5em minus 0.4em\relax Boston, MA: Springer US, 2011, pp. 1--35.

\bibitem{lupton17}
D.~Lupton and B.~Williamson, ``The datafied child: The dataveillance of children and implications for their rights,'' \emph{New Media \& Society}, vol.~19, no.~5, pp. 780--794, 2017.

\end{thebibliography}

\end{document}